\documentclass[10pt,a4paper]{article}

\usepackage{graphics,tikz,amsmath}

\newcommand{\bea}{\begin{eqnarray}}
\newcommand{\eea}{\end{eqnarray}}
\newcommand{\be}{\begin{equation}}
\newcommand{\ee}{\end{equation}}

\def\be{\begin{eqnarray}}
\def\ee{\end{eqnarray}}
\def\bd{\begin{displaymath}}
\def\ed{\end{displaymath}}

\def\ga{\gamma}
\def\etal{{\em et al}}
\def\ADNDT{{\it At. Data. Nucl. Data. Tables }}

\def\PRL{Phys. Rev. Lett. }

\def\jpg{{\it J. Phys. G: Nucl. Part. Phys.}}

\begin{document}
\title{Proton dripline in a new formula for nuclear binding energy}
\author{Chirashree Lahiri and G. Gangopadhyay\\
Department of Physics, University of Calcutta\\
92, Acharya Prafulla Chandra Road, Kolkata-700 009, India\\
email: ggphy@caluniv.ac.in}
\maketitle

\begin{abstract}
The location of the proton dripline in a new phenomenological
mass formula is calculated. Predictions of different mass formulas for
the dripline are compared. The implications of the new mass formula for 
rapid proton nucleosynthesis beyond $^{56}$Ni are discussed.
It is seen that the new formula indicates that masses up to $A=80$ are 
easily synthesized in a typical X-ray burst.
\end{abstract}

\section{Introduction}

Prediction of binding energy is a very important task in  nuclear physics.
For example, binding energy of the nuclear ground state is one of the most 
important inputs in the study of astrophysical reactions. 
Experimental mass measurements are difficult to make in nuclei far from the 
stability valley. Thus, one has to take recourse to theoretical predictions. 
In a recent paper\cite{massform} (hereafter called Ref. I) a new 
phenomenological formula for ground state binding energies has been introduced. In the present 
work, we explore  a few of the implications of the new formula and also
compare our results with the predictions from some other formulas.

The area we concentrate on in the present work is related to the study of 
proton dripline. As pointed out in Ref. I, the root men square (r.m.s.) error 
in the ground state binding energies is 0.376 MeV for 2140 nuclei. 
Near the stability valley, single nucleon separation energy (proton or 
neutron) is of the order of 8 MeV. Thus, a small error in binding energy
does not influence the results of processes which involve nuclei near
this valley. However, the separation energies of protons and neutrons near 
the corresponding driplines are actually very small, sometimes comparable to 
the errors in theoretical predictions. Thus, it should be interesting to 
study the performance of the formula from Ref. I in predicting the location 
of the proton dripline. We also investigate the effect of the mass formula on the
rapid proton ($rp$) process nucleosynthesis beyond $^{56}$Ni. This process
goes along the $N=Z$ line and is dependent on the binding energy of protons
near the proton dripline.

\section{Calculation and Results}

For details of the mass formula, we refer the readers to Ref I where it was 
developed. The present section is divided into two parts. In the first part, we 
calculate the location of the proton dripline according to the present formula
and compare with experimental observations and a few other predictions. In the 
second part, we study the
effect of the present mass formula in astrophysical $rp$-process
in mass 60-80 region.

\subsection{Proton dripline}

The proton dripline for a particular neutron number is defined to be the 
nucleus beyond which the proton separation energy becomes zero or negative. 
Obviously, because of the 
pairing effect, the $Z$ of the proton dripline nucleus is expected to be even.
Studying the nucleon dripline is one of the major activities in nuclear 
physics. However, it has been possible to reach the neutron dripline in only 
very light nuclei. Proton dripline, on the other hand, is more accessible
to experiments. We note that the location of the proton dripline is known for a 
number of neutron numbers, particularly in mass 100-220 region. Nuclei with
protons in positive energy continuum show proton radioactivity where protons
tunnel through the Coulomb barrier. A number of such nuclei are known in 
mass 100-200 region. 

\begin{table}[bht]
\caption{Location of proton dripline. The columns stand for the following:
Exp. - Experimental, Ref. I - Present calculation, FRDM - Finite Range  
Droplet Model and D-Z - Duflo-Zuker formula.}
\begin{tabular}{rcccc|rcccc}\hline
N&\multicolumn{4}{c|}{$Z$ of proton dripline nucleus}&
N&\multicolumn{4}{c}{$Z$ of proton dripline nucleus}\\
 &\multicolumn{1}{c}{Exp.} &\multicolumn{1}{c}{Ref. I}  
&\multicolumn{1}{c}{FRDM} &\multicolumn{1}{c|}{D-Z}&
 &\multicolumn{1}{c}{Exp.} &\multicolumn{1}{c}{Ref. I}  
&\multicolumn{1}{c}{FRDM} &\multicolumn{1}{c}{D-Z}
\\\hline
18 & 20&20&20&20& 
96&78&78&78&78\\
32 & 32& 32&34&32& 
97&78&80&78&78\\
34 & 34&34&36&34& 
98&78&80&78&80\\
36 & 36 & 36&36&36& 
99&80&80&78&80\\
54 & 50&50&50&50& 
100&80&80&80&80\\
56&52&52&50&52& 
101&80&80&82&82\\
58&54&54&54&54& 
102&82&80&82&82\\
78&66&68&68&68& 
103&82&82&82&82\\
79&68&68&70&68& 
104&82&82&82&82\\
82&70&70&70&70& 
105&82&82&82&82\\
83&70&72&72&70& 
106&82&82&82&82\\
84&70&72&72&72& 
108&84&84&82&84\\
85&72&72&72&72& 
110&84&84&86&84\\
86&72&72&74&72& 
112&86&86&86&86\\
87&74&74&74&72& 
113&86&86&86&86\\
88&74&74&74&74& 
114&86&86&88&86\\
90&74&74&76&74& 
117&88&88&88&88\\
91&76&76&76&76& 
118&88&88&88&88\\
92&76&76&76&76& 
121&90&90&90&90\\
93&76&78&78&76& 
122&90&90&90&90\\
94&76&78&78&76& 
123&90&92&92&92\\
\hline
\end{tabular}
\end{table}
 
The known mass values of nuclei enable us to calculate the location of the
proton dripline. In the first two columns of table I, we summarize our 
knowledge for the neutron numbers for which the experimental proton dripline 
is exactly known. The mass table of Audi \etal.\cite{audi} and the more 
recent references on binding energy measurements included in Ref. I have been used to determine 
the experimental proton dripline. We calculate the location of the proton 
dripline from the new binding energy formula of Ref. I and present it in Table 
I. The results of the Finite Range Droplet Model (FRDM)\cite{Moller} and 
the Duflo-Zuker (D-Z) mass formula\cite{DZ} 
are also presented. We see that the present method can accurately 
predict the location of the proton dripline in most of the cases and is actually
better than FRDM. The Duflo-Zuker formula is even better in this respect.

We would like to point out that the conclusions of different experiments on the 
binding energy of $^{65}$As do not agree. Schury \etal.\cite{prc} have 
concluded that $^{65}$As has a negative proton separation energy while a 
recent measurement\cite{csb} has raised the possibility of it being stable
with respect to proton emission. The implication of this uncertainty in proton 
separation energy of $^{65}$As has been discussed in the next subsection in 
more detail.

Even in the few cases where the present approach fails to accurately locate the 
dripline, the next even $Z$ nuclei is indicated. One needs to remember that
the formula has an r.m.s. error of 376 keV. Thus, if the predicted binding 
energy of the last proton in an odd $Z$ nucleus is very small, it is possible 
that the nucleus is actually beyond the proton dripline. Conversely, prediction
of a very small negative value of last proton separation energy cannot guarantee
that the dripline does not actually lie beyond.
 For example, in $N=93$ and $94$ nuclei, the calculated binding energy of 
the last proton in $_{77}^{170,171}$Lu are only 0.047 MeV and 0.012 MeV, 
respectively. On the other hand, the calculated separation energy of the last 
proton in $_{81}^{183}$Tl ($N=102$) is -0.030 MeV only.

\subsection{Astrophysical rapid proton process}

We want to look at the effect of the new mass formula on nucleosynthesis. 
Particularly, in view of the success of the formula to predict the proton 
dripline, it should be interesting to see the effect of the proton separation 
energy predicted by the present formula on the 
$rp$-process. In this process, which takes place in hot 
explosive proton-rich environment such as X-ray bursts, a nucleus may capture 
high energy protons. However, this process has to compete with its inverse, 
 {\rm i.e.} photodisintegration by emitting a proton at high temperature
(($\ga,p$) reaction).
A negative or a small positive value of proton separation energy implies that
the inverse reaction dominates and the $rp$-process stalls at that point, 
the so-called waiting point. Two proton capture may lead to the 
nucleosynthesis bridging the waiting point and proceed beyond.
More details of the process is available in standard text books (For example 
the book by Illiadis\cite{book}).

The stellar decay constant ($\lambda$) for  photodisintegration, {\em i.e.} 
($\ga,p$) 
reaction, is related to the proton capture rate by the reciprocity theorem in 
the following numerical form\cite{book}
\begin{eqnarray}
\lambda=
9.86851\times10^9 T^{\frac{3}{2}}\left(\frac{M_{p}M_{X}}{M_{Y}}\right)^{\frac{3}{2}}
\frac{\left( 2J_{p}+1 \right)\left( 2J_{X}+1 \right)}
{\left( 2J_{Y}+1 \right)}\frac{G_pG_X}{G_Y}
\nonumber\\
{N\langle\sigma v\rangle_{pX\rightarrow Y\ga}} 
\exp\left(\frac{-11.605Q}{T}\right)
\end{eqnarray}
for the reaction $p+X\longleftrightarrow Y+\gamma$ and is expressed in 
$sec^{-1}$ when the forward reaction rate, given by 
$N\langle\sigma v\rangle_{pX\rightarrow Y\ga}$, is expressed in 
$cm^3mol^{-1}sec^{-1}$. The 
temperature $T$ is expressed in GK ($10^9$K) and $Q$ is the ground state Q-value of the 
forward reaction expressed in MeV. The normalized partition functions, $G_X$
and $G_Y$, 
may be obtained from Rauscher \etal.\cite{ptable}. For protons, we have $G_p=1$
and $J_p=1/2$.

The photodisintegration rate depends exponentially on the  proton separation 
energy. This quantity, thus, plays a very important role in the possibility of 
$rp$-process proceeding beyond the waiting point. There are significant 
disagreements in the predictions of proton separation energy of different 
mass formulas, and even between different measurements. 
In some other cases, experimental measurements may exist but the the error in 
measurement is too large to draw any meaningful conclusion. As an example,
we look at an example of the waiting point nucleus, $^{64}$Ge. The 
measured and estimated
Q-values of the proton capture reaction differ over a large range. Schury 
\etal.\cite{prc} have measured the Q-value to be $-0.255\pm 0.104$ MeV.
A recent measurement\cite{csb} has found the proton separation energy of 
$^{65}$As to be 0.401(530) MeV, {\em i.e.} the nucleus may even be bound.
The Duflo-Zuker formula\cite{DZ} suggests the value to be -0.407 MeV. 
The FRDM calculation of M\"{o}ller \etal.\cite{Moller} predicts the value as 
0.13 MeV. 
The mass table\cite{audi} gives the Q-value as -0.08 MeV. A very recent 
measurement\cite{Asnew} has given a value of -0.090(0.085) MeV.
The new mass formula suggests a value -0.116 MeV, which is very close to the last 
estimation. In such cases, it becomes imperative to look at the implications 
of different mass formulas. We present the bridging this
particular waiting point in some detail.

For the calculation of proton capture process, a small network  has been
designed which includes the following processes. The waiting point nucleus, 
which acts as a seed, may capture a proton. The resulting nucleus $^{65}$As
may either capture another proton to produce $^{66}$Se or undergo 
photodisintegration emitting a proton to go back to the seed nucleus. The 
nucleus $^{66}$Se  may also undergo photodisintegration. In addition, all the 
three nuclei mentioned above may undergo $\beta$-decay. 

We calculate the change in half life of $^{64}$Ge in an explosive 
astrophysical environment assuming the density to be $10^6$gm/cm$^3$ and the 
proton fraction to be 0.7. A microscopic optical potential has been obtained by 
folding the DDM3Y potential\cite{ddm3y} with nuclear densities obtained in 
Relativistic Mean Field model with the Lagrangian density FSU Gold\cite{fsu}. 
The calculation follows our earlier works\cite{cl} where more details are available.
The measured half life values for $\beta$-decay have been taken from
the compilation by Audi \etal.\cite{Audi1} except in the case of
the case of $^{65}$As. For this nucleus, a value of 0.128
seconds from an experimental measurement\cite{as65} has been assumed.
If measured values are not available, we have adopted the values from the
work by M\"{o}ller \etal.\cite{Moller1}

\begin{figure}[htb]
\center
\resizebox{6.5cm}{!}{\includegraphics{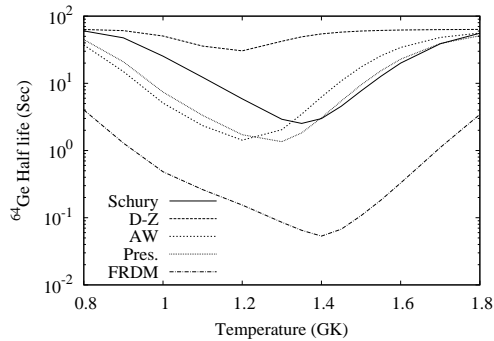}}\hskip -0.5cm
\caption{\label{ge64}Effective half life values of $^{64}$Ge as a function of
temperature for different Q-values of $^{64}$Ge$(p,\ga)$ reaction.} 

\end{figure}

The results of the calculation on effective half life of $^{64}$Ge as a 
function of temperature are presented  for different mass estimates in
Fig. 1. The different values have been indicated as follows: Schury--
measurement by Schury \etal.\cite{prc}, D-Z-- Duflo-Zuker formula\cite{DZ}, 
A-W-- Mass table of Audi \etal.\cite{audi}, Pres.-- Present work and
FRDM-- FRDM results of M\"{o}ller \etal.\cite{Moller}.  
At low temperature proton capture rate is small and 
$\beta$-decay
predominates. At high temperature photodisintegration dominates over capture.
Thus the predominant decay mode is again $\beta$-decay. In between there is a 
temperature window where two proton capture leads to $^{66}$Se.
From the figure, it is clear that the rate of two proton capture is 
sensitively dependent on the masses used in the calculation. The Q-values
predicted by the Duflo-Zuker formula leads to a minimum lifetime of 
nearly 30 seconds. The values predicted by the present mass formula and 
that by  Audi \etal.\cite{audi} decrease the lifetime by more than one order of 
magnitude.  If $^{65}$As is bound as suggested by some works\cite{csb,Moller}, 
the rate of the inverse process is small. In such a situation, two 
proton capture dominates and the waiting point is easily bridged.

As is evident from the above discussion, the bridging of waiting points, and 
hence the $rp$-process abundance crucially depends on the proton separation 
energy. However, this is sometimes poorly known on the $N=Z$ nucleus as the 
process goes along the proton dripline where information is meager. The degree 
of bridging of the waiting point has significant effect
on the path that the $rp$-nucleosynthesis takes in explosive 
proton-rich environments such as X-ray bursters. Typical X-ray burst 
has a time scale of 10-100 seconds. In case a significant fraction
of the population does not undergo two proton capture reaction, 
it traverses along a path of more stable isotopes and nucleosynthesis
is delayed. Thus, it may not be possible for an X-ray burst to
populate nuclei with significant mass in case a particular waiting point stalls 
the process substantially.

\begin{figure}[htb]
\resizebox{6.5cm}{5cm}{\includegraphics{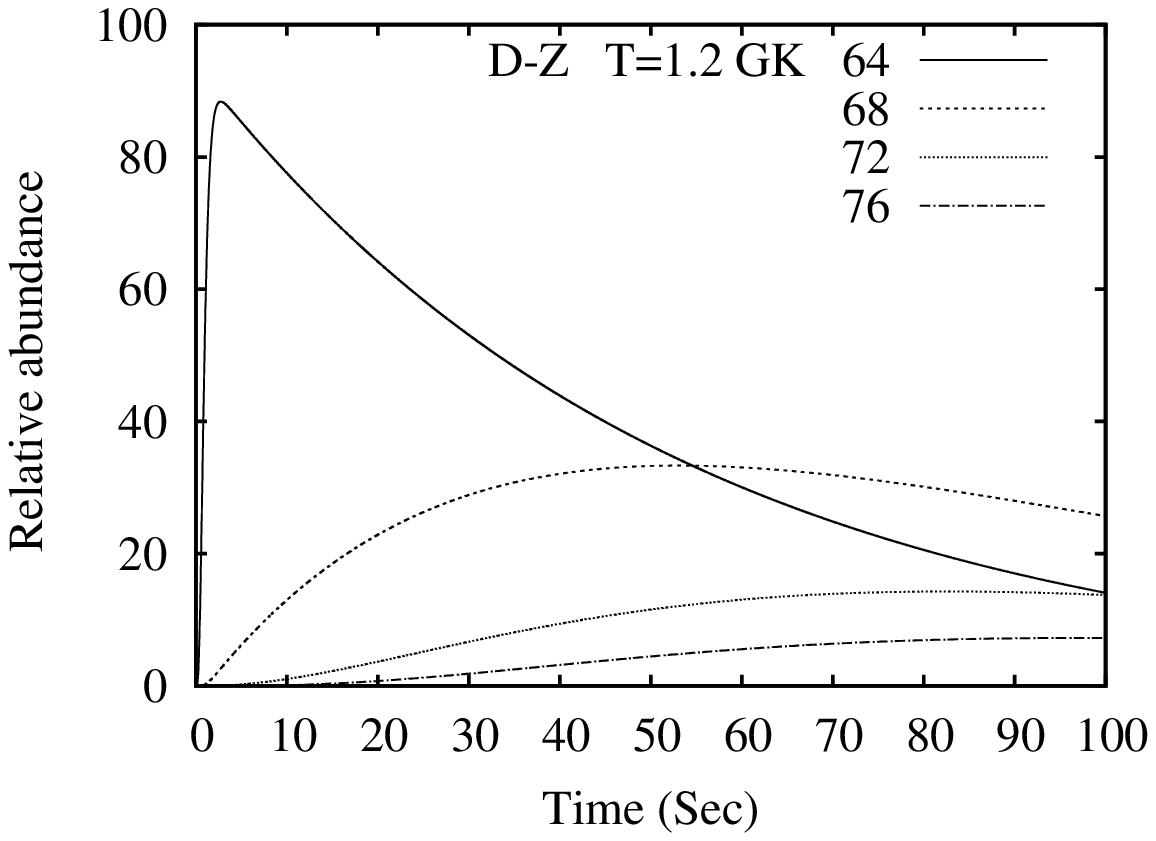}}\hskip -0.5cm 
\resizebox{6.5cm}{5cm}{\includegraphics{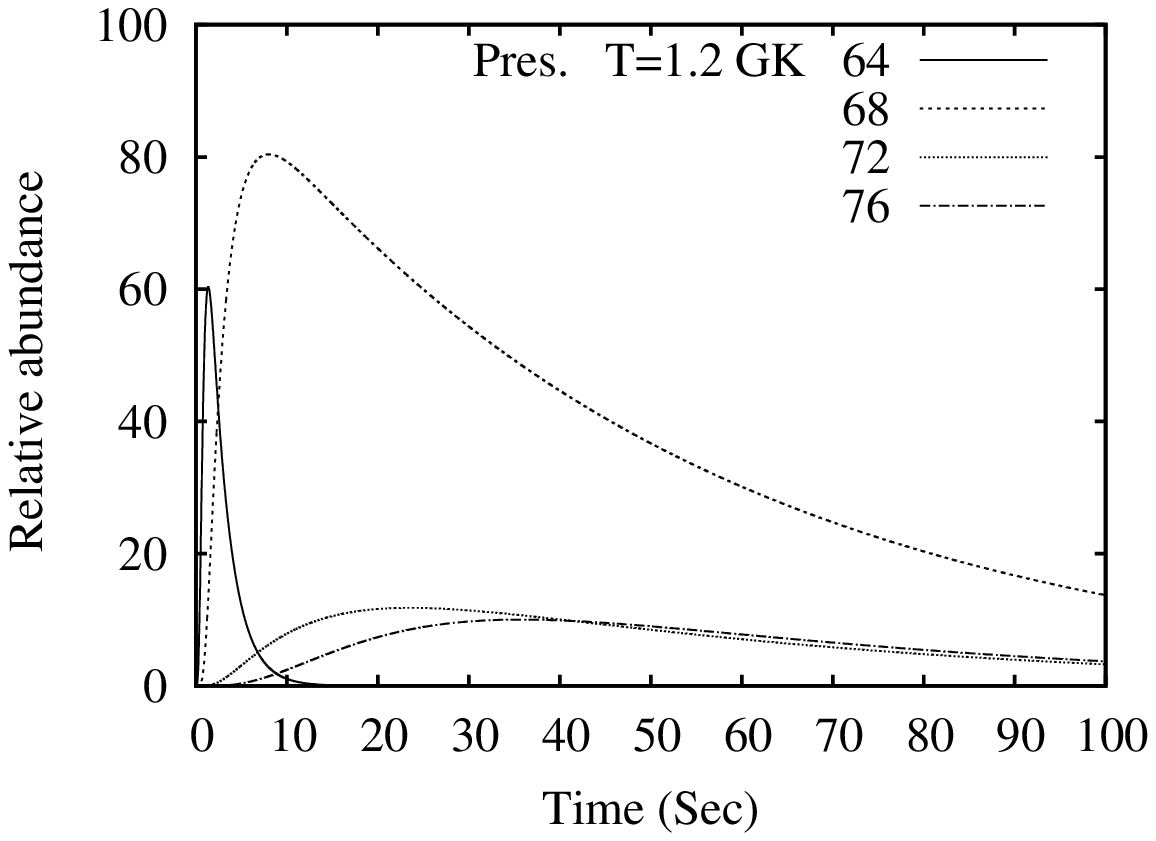}} 
\resizebox{6.635cm}{5cm}{\includegraphics{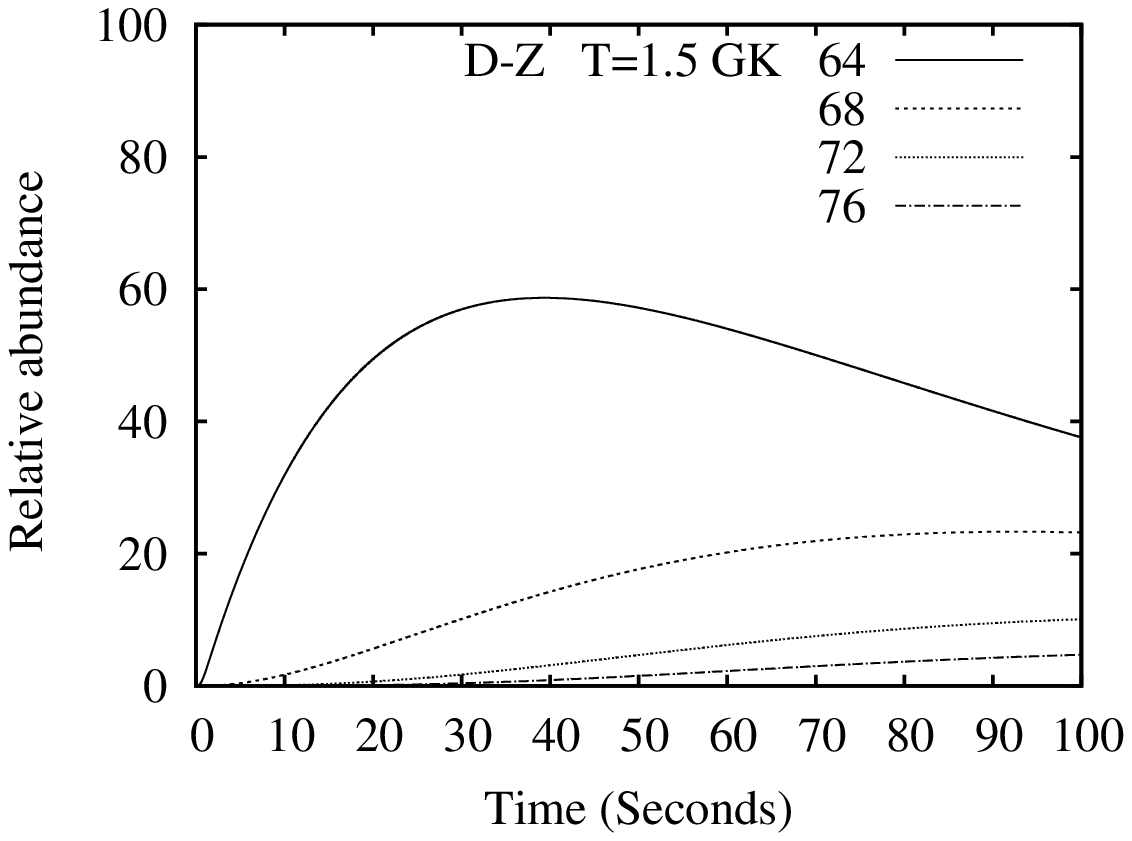}}\hskip -0.635cm 
\resizebox{6.635cm}{5cm}{\includegraphics{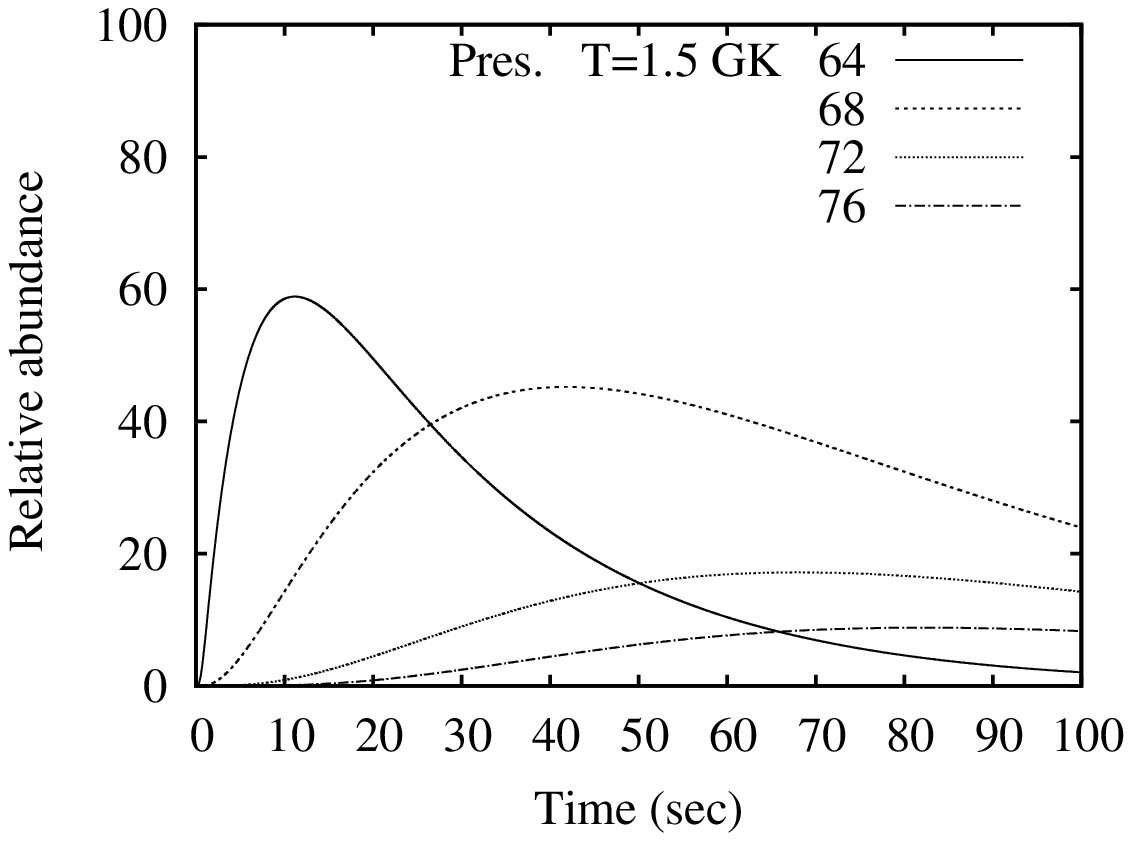}} 
\caption{Evolution of abundance of mass at the waiting points in explosive
proton-rich astrophysical environments. See text for details.\label{net}}
\end{figure}

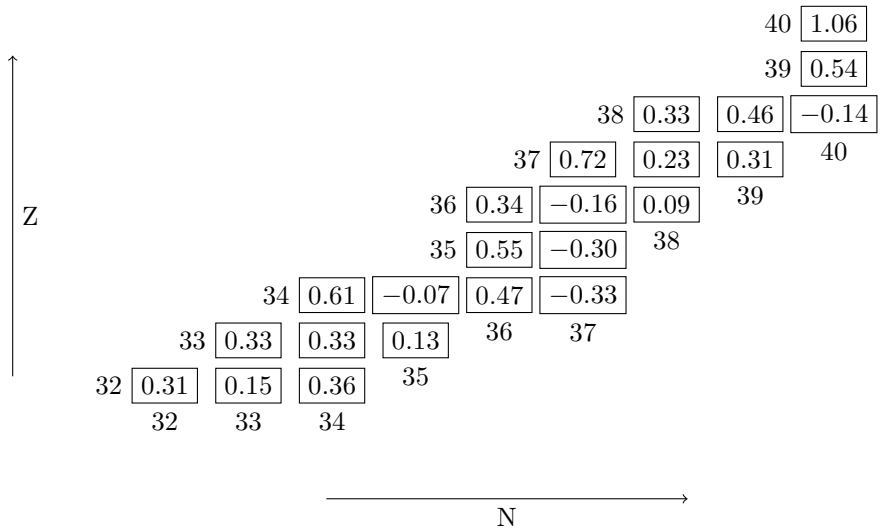
\begin{figure}[hbt]
\begin{tikzpicture}
\node (x1) at (3,-.5)  {};
\node (x2) at (8,-.5)  {};
\draw [->] (x1.east) to node [auto,swap] {N} (x2.west);
 \node (y1) at (-1,1)  {};
\node (y2) at (-1,5.5)  {};
\draw [->] (y1.north) to node [auto,swap] {Z} (y2.south);

\node (1) at (1,1) [draw,label=left:32,label=below:32] {$0.31$};
 \node (2) at (2.1,1) [draw,label=below:33] {$0.15 $};
\node (3) at (3.2,1) [draw,label=below:34] {$0.36 $};
\node (5) at (2.1,1.6) [draw,label=left:33] {$0.33$};
\node (6) at (3.2,1.6) [draw] {$0.33$};
\node (7) at (4.3,1.6) [draw,label=below:35] {$0.13$};
\node (8) at (3.2,2.2) [draw,label=left:{34}] {$0.61$};
\node (9) at (4.3,2.2) [draw] {$-0.07$};
\node (10) at (5.4,2.2) [draw,label=below:36] {$0.47$};
\node (11) at (6.5,2.2) [draw,label=below:37] {$-0.33$};
\node (13) at (5.4,2.8) [draw,label=left:35] {$0.55$};
\node (14) at (6.5,2.8) [draw] {$-0.30$};
\node (15) at (5.4,3.4) [draw,label=left:36] {$0.34$};
\node (16) at (6.5,3.4) [draw] {$-0.16$};
\node (17) at (7.6,3.4) [draw,label=below:38] {$0.09$};
\node (18) at (6.5,4.) [draw,label=left:37] {$0.72$};
\node (19) at (7.6,4) [draw] {$0.23$};
\node (20) at (8.7,4) [draw,label=below:39] {$0.31$};
\node (21) at (7.6,4.6) [draw,label=left:38] {$0.33$};
\node (22) at (8.7,4.6) [draw] {$0.46$};
\node (23) at (9.8,4.6) [draw,label=below:40] {$-0.14$};
\node (25) at (9.8,5.2) [draw,label=left:39] {$0.54$};
\node (26) at (9.8,5.8) [draw,label=left:40] {$1.06$};
\end{tikzpicture}
\caption{\label{mass} Differences between experimental\cite{audi} and 
calculated mass values near the $N=Z$ line in mass 60-80 region. }
\end{figure} 

As an example, let us compare the effects of the Duflo-Zuker and the new mass 
formulas on the population at different masses compared to the waiting points 
with $A<80$. We have calculated the $rp$-process rates for mass 60-80 region. 
We plot the difference of the calculated binding energy using Ref I from
the experimental values in Fig. 2.
The mass formulas have been used for masses which have not been 
experimentally measured or have considerable experimental uncertainty as in the 
case of $^{65}$As. For consistency, even if one of the masses involved in the 
reaction is known, we have obtained the Q-values from theoretical predictions 
only. The waiting points occur at A=64, 68, 72 and 76 in the nuclei with $N=Z$. 
We assume a constant proton fraction of 0.7. 
We have designed a network which includes the proton-capture reactions and 
$\beta$-decay starting from the seed nucleus $^{56}$Ni and goes beyond $A=80$. 
Astrophysical rates have been calculated as discussed in our previous 
works\cite{cl}. The 
network is run for two constant temperatures, 1.2 GK and 1.5 GK.
In Fig \ref{net}, we plot the evolution of abundance of nuclear mass as a 
function of time for $A=64$, 68, 72 and 76. The top (bottom) two panels are the results for $T=1.2$ GK 
(1.5 GK). The left (right) panels indicate the results obtained when we use 
the Duflo-Zuker formula (Ref I) for unknown mass values as described above.

It is clear the Duflo-Zuker mass formula predicts a significant stalling at 
mass $A=64$ so much so that a burst of ten second duration fails to 
proceed significantly beyond this mass. It is consistent with the result on 
half life discussed earlier as at 1.5 GK, the half life of $^{64}$Ge is not 
expected to change significantly. Thus the $rp$-process can continue only after $\beta$-decay of $^{64}$Ge.
The process thus shifts toward more stable nuclei and gets stalled.
We see that the abundance is dominated my $A=64$ nuclei the rapid proton 
nucleosynthesis even in an X-ray burst of 100 seconds for $T=1.5$ GK.
 
The present mass formula, on the other hand, indicates that most of the
$^{64}$Ge is converted to $^{66}$Se very shortly.
This leads to significant differences in the abundance pattern.
Of course, in an actual X-ray burst, the temperature  and the proton mass 
fraction are not constant but change with time. Thus we do not expect this 
simple calculation to reproduce the observed abundance pattern. The aim of the
present study is to point out the effect the different mass formulas can have 
on the final abundance pattern. 

\begin{figure}[hbt]
\begin{tikzpicture}
\node (x1) at (0,-3)  {};
\node (x2) at (6,-3)  {};

  \node (1) at (.5,0)  [draw,inner sep=7.1pt,label=left:Ni(28),label=below :28] {$ $  };
    \node (2) at (.5,.5) [draw,inner sep=7.1pt,label=left:Cu(29)] {$ $};
    \node (3) at (.5,1) [draw,inner sep=7.1pt,label=left:Zn(30)] {$ $};
    \node (4) at (.5,1.5) [draw,inner sep=7.1pt,label=left:Ga(31)] {$ $};
    \node (5) at (1,.5) [draw,inner sep=7.1pt]{$ $};
    \node (6) at (1.,1) [draw,inner sep=7.1pt]{$ $};
 \node (7) at (1.,1.5) [draw,inner sep=7.1pt]   {$ $ };
 \node (8) at (1.5,.5) [draw,inner sep=7.1pt]    {$ $};
 \node (9) at (1.5,1) [draw,inner sep=7.1pt]    {$ $};
 \node (10) at (1.5,1.5) [draw,inner sep=7.1pt]   {$ $};
\node (11) at (1.5,2.) [draw,inner sep=7.1pt,label=left:Ge(32)]    {$ $};
 \node (12) at (2.,.5) [draw,inner sep=7.1pt] {$ $};
 \node (13) at (2.0,1) [draw,inner sep=7.1pt] {$ $};
 \node (14) at (2.,1.5) [draw,inner sep=7.1pt] {$ $};
\node (15) at (2.,2.) [draw,inner sep=7.1pt]  {$ $};
 \node (16) at (2.5,1.5) [draw,inner sep=7.1pt]   {$ $};
\node (17) at (2.5,2.) [fill,draw,inner sep=7.1pt]    {$ $};
 \node (18) at (2.5,2.5) [draw,inner sep=7.1pt]   {$ $};
\node (19) at (2.5,3.) [draw,inner sep=7.1pt]    {$ $};
 \node (20) at (3.,1.5) [draw,inner sep=7.1pt] {$ $};
 \node (21) at (1.0,0) [draw,inner sep=7.1pt]  {$ $};
 \node (22) at (1.5,0) [draw,inner sep=7.1pt,label=below :30]   {$ $};
 \node (23) at (2.0,0) [draw,inner sep=7.1pt] {$ $};
 \node (24) at (2.5,0) [draw,inner sep=7.1pt,label=below :32]   {$ $};
 \node (25) at (2.5,.5) [draw,inner sep=7.1pt]   {$ $};
 \node (26) at (2.5,1) [draw,inner sep=7.1pt]   {$ $};
\node (27) at (1.5,2.5) [draw,inner sep=7.1pt,label=left:As(33)] {$ $};
\node (28) at (1.5,3.) [draw,inner sep=7.1pt,label=left:Se(34)] {$ $};
\node (29) at (2.,2.5) [draw,inner sep=7.1pt] {$ $};
\node (30) at (2.,3.) [draw,inner sep=7.1pt] {$ $};
\node (31) at (3.,2.) [draw,inner sep=7.1pt] {$ $};
\node (32) at (3.,2.5) [draw,inner sep=7.1pt] {$ $};
\node (33) at (3.,3.) [draw,inner sep=7.1pt] {$ $};
\node (34) at (3.,3.5) [draw,inner sep=7.1pt,label=left:Br(35)] {$ $};
 \node (35) at (3.5,1) [draw,inner sep=7.1pt,label=below :34] {$ $};
 \node (36) at (3.5,1.5) [draw,inner sep=7.1pt] {$ $};
 \node (37) at (3.5,2.) [draw,inner sep=7.1pt] {$ $};
\node (38) at (3.5,2.5) [draw,inner sep=7.1pt] {$ $};
\node (39) at (3.5,3) [fill,draw,inner sep=7.1pt] {$ $};
\node (40) at (3.5,3.5) [draw,inner sep=7.1pt] {$ $};
\node (41) at (3.5,4.) [draw,inner sep=7.1pt,label=left:Kr(36)] {$ $};
\node (42) at (4.,2.5) [draw,inner sep=7.1pt] {$ $};
\node (43) at (4.,3.) [draw,inner sep=7.1pt] {$ $};
\node (44) at (4.0,3.5) [draw,inner sep=7.1pt] {$ $};
\node (45) at (4.0,4.) [draw,inner sep=7.1pt] {$ $};
\node (46) at (4.5,3.) [draw,inner sep=7.1pt,label=below:36] {$ $ };
\node (47) at (4.5,3.5) [draw,inner sep=7.1pt] {$ $ };
\node (48) at (4.5,4.) [fill,draw,inner sep=7.1pt] {$ $};
\node (49) at (4.5,4.5) [draw,inner sep=7.1pt,label=left:Rb(37)] {$ $ };
\node (50) at (4.5,5.) [draw,inner sep=7.1pt,label=left:Sr(38)] {$ $ };
\node (51) at (5.,3.5) [draw,inner sep=7.1pt,label=below:37] {$ $ };
\node (52) at (5,4.) [draw,inner sep=7.15pt] {$ $  };
\node (53) at (5.,4.5) [draw,inner sep=7.1pt] {$ $ };
\node (54) at (5.,5.) [draw,inner sep=7.1pt] {$   $ };
\node (55) at (5.,5.5) [draw,inner sep=7.1pt,label=left:Y(39)] {$  $ };
\node (56) at (5.5,4.) [draw,inner sep=7.1pt,label=below:38] {$ $ };
\node (57) at (5.5,4.5) [draw,inner sep=7.1pt] {$  $ };
\node (58) at (5.5,5) [fill,draw,inner sep=7.1pt] {$ $ };
\node (59) at (5.5,5.5) [draw,inner sep=7.1pt] {$ $ };
\node (60) at (5.5,6.) [draw,inner sep=7.1pt,label=left:Zr(40)] {$ $ };
\node (61) at (6.,4.5) [draw,inner sep=7.1pt,label=below:39] {$ $ };
\node (62) at (6.,5.) [draw,inner sep=7.1pt] {$ $ };
\node (63) at (6.,5.5) [draw,inner sep=7.1pt] {$ $ };
\node (64) at (6.,6.) [draw,inner sep=7.1pt] {$ $ };
\node (66) at (6.5,5.) [draw,inner sep=7.1pt,label=below:40] {$ $ };
\node (67) at (6.5,5.5) [draw,inner sep=7.1pt] {$ $ };
\node (68) at (6.5,6.) [fill,draw,inner sep=7.1pt] {$ $ };
\node (71) at (3.,1.) [draw,inner sep=7.1pt] {$ $ };

\draw [very thick]  (.5,0) to (.5,.5); 
\draw [very thick]  (.5,.5) to (.5,1.); 
\draw [very thick]  (.5,1.) to (1.0,.5); 
\draw [very thick]  (1.0,.5) to (1.0,1.); 
\draw [very thick]  (1.0,1.) to (1.5,.5); 
\draw [very thick]  (1.5,.5) to (1.5,1.); 
\draw [very thick]  (1.5,1) to (1.5,1.5); 
\draw [very thick]  (1.5,1.5) to (1.5,2.); 
\draw [very thick]  (1.5,2.) to (2.,1.5);
\draw [very thick]  (2.,1.5) to (2.,2.);
\draw [very thick]  (2.,2.) to (2.5,1.5); 
\draw [very thick]  (2.5,1.5) to (2.5,2.);
\draw [very thick]  (2.5,2.) to (2.5,2.5);
\draw [thick, gray]  (2.5,2.) to (3.0,1.5);
\draw [thick, gray]  (3.0,1.5) to (3.0,2.5);
\draw [very thick]  (2.5,2.5) to (2.5,3.);
\draw [very thick]  (2.5,3.) to (3.,2.5);
\draw [very thick]  (3.,2.5) to (3.,3.);
\draw [very thick]  (3.,3.) to (3.5,2.5);
\draw [very thick]  (3.5,2.5) to (3.5,3.);
\draw [very thick]  (4.,3.5) to (4.,4.);
\draw [very thick]  (4.,4.) to (4.5,3.5);
\draw [very thick]  (4.5,3.5) to (4.5,4.);
\draw [very thick]  (4.5,4.) to (4.5,4.5);
\draw [very thick]  (4.5,4.5) to (4.5,5.);
\draw [very thick]  (4.5,5) to (5,4.5);
\draw [very thick]  (5.,4.5) to (5,5.);
\draw [very thick]  (5,5.) to (5.5,4.5);
\draw [very thick] (5.5,4.5) to (5.5,5.);
\draw [very thick] (5.5,5) to (6,4.5);
\draw [very thick] (6,4.5) to (6,5.5);
\draw [thick, gray] (5.5,5) to (5.5,5.5);
\draw [thick, gray] (5.5,5.5) to (5.5,6.);
\draw [thick, gray] (5.5,6.) to (6,5.5);
\draw [very thick] (6,5.5) to (6.,6.);
\draw [very thick] (6.,6.) to (6.5,5.5);
\draw [very thick] (6.5,5.5) to (6.5,6.);
\draw [very thick] (4.5,4.) to (5,3.5);
\draw [very thick] (5,3.5) to (5,4.);
\draw [very thick] (5,4.) to (5,4.5);
\draw [very thick] (5,4.5) to (5,5.);
\draw [very thick] (3.5,3.) to (4.,2.5);
\draw [very thick] (4.,2.5) to (4.,3.);
\draw [very thick] (4.,3.) to (4.,3.5);

\end{tikzpicture}
\vskip -8cm
\begin{tikzpicture}
\node (x1) at (-5.5,-3)  {};
\node (x2) at (0.5,-3)  {};

  \node (1) at (.5,0)  [draw,inner sep=7.1pt,label=left:Ni(28),label=below :28] {$ $  };
    \node (2) at (.5,.5) [draw,inner sep=7.1pt,label=left:Cu(29)] {$ $};
    \node (3) at (.5,1) [draw,inner sep=7.1pt,label=left:Zn(30)] {$ $};
    \node (4) at (.5,1.5) [draw,inner sep=7.1pt,label=left:Ga(31)] {$ $};
    \node (5) at (1,.5) [draw,inner sep=7.1pt]{$ $};
    \node (6) at (1.,1) [draw,inner sep=7.1pt]{$ $};
 \node (7) at (1.,1.5) [draw,inner sep=7.1pt]   {$ $ };
 \node (8) at (1.5,.5) [draw,inner sep=7.1pt]    {$ $};
 \node (9) at (1.5,1) [draw,inner sep=7.1pt]    {$ $};
 \node (10) at (1.5,1.5) [draw,inner sep=7.1pt]   {$ $};
\node (11) at (1.5,2.) [draw,inner sep=7.1pt,label=left:Ge(32)]    {$ $};
 \node (12) at (2.,.5) [draw,inner sep=7.1pt] {$ $};
 \node (13) at (2.0,1) [draw,inner sep=7.1pt] {$ $};
 \node (14) at (2.,1.5) [draw,inner sep=7.1pt] {$ $};
\node (15) at (2.,2.) [draw,inner sep=7.1pt]  {$ $};
 \node (16) at (2.5,1.5) [draw,inner sep=7.1pt]   {$ $};
\node (17) at (2.5,2.) [fill,draw,inner sep=7.1pt]    {$ $};
 \node (18) at (2.5,2.5) [draw,inner sep=7.1pt]   {$ $};
\node (19) at (2.5,3.) [draw,inner sep=7.1pt]    {$ $};
 \node (20) at (3.,1.5) [draw,inner sep=7.1pt] {$ $};
 \node (21) at (1.0,0) [draw,inner sep=7.1pt]  {$ $};
 \node (22) at (1.5,0) [draw,inner sep=7.1pt,label=below :30]   {$ $};
 \node (23) at (2.0,0) [draw,inner sep=7.1pt] {$ $};
 \node (24) at (2.5,0) [draw,inner sep=7.1pt,label=below :32]   {$ $};
 \node (25) at (2.5,.5) [draw,inner sep=7.1pt]   {$ $};
 \node (26) at (2.5,1) [draw,inner sep=7.1pt]   {$ $};
\node (27) at (1.5,2.5) [draw,inner sep=7.1pt,label=left:As(33)] {$ $};
\node (28) at (1.5,3.) [draw,inner sep=7.1pt,label=left:Se(34)] {$ $};
\node (29) at (2.,2.5) [draw,inner sep=7.1pt] {$ $};
\node (30) at (2.,3.) [draw,inner sep=7.1pt] {$ $};
\node (31) at (3.,2.) [draw,inner sep=7.1pt] {$ $};
\node (32) at (3.,2.5) [draw,inner sep=7.1pt] {$ $};
\node (33) at (3.,3.) [draw,inner sep=7.1pt] {$ $};
\node (34) at (3.,3.5) [draw,inner sep=7.1pt,label=left:Br(35)] {$ $};
 \node (35) at (3.5,1) [draw,inner sep=7.1pt,label=below :34] {$ $};
 \node (36) at (3.5,1.5) [draw,inner sep=7.1pt] {$ $};
 \node (37) at (3.5,2.) [draw,inner sep=7.1pt] {$ $};
\node (38) at (3.5,2.5) [draw,inner sep=7.1pt] {$ $};
\node (39) at (3.5,3) [fill,draw,inner sep=7.1pt] {$ $};
\node (40) at (3.5,3.5) [draw,inner sep=7.1pt] {$ $};
\node (41) at (3.5,4.) [draw,inner sep=7.1pt,label=left:Kr(36)] {$ $};
\node (42) at (4.,2.5) [draw,inner sep=7.1pt] {$ $};
\node (43) at (4.,3.) [draw,inner sep=7.1pt] {$ $};
\node (44) at (4.0,3.5) [draw,inner sep=7.1pt] {$ $};
\node (45) at (4.0,4.) [draw,inner sep=7.1pt] {$ $};
\node (46) at (4.5,3.) [draw,inner sep=7.1pt,label=below:36] {$ $ };
\node (47) at (4.5,3.5) [draw,inner sep=7.1pt] {$ $ };
\node (48) at (4.5,4.) [fill,draw,inner sep=7.1pt] {$ $};
\node (49) at (4.5,4.5) [draw,inner sep=7.1pt,label=left:Rb(37)] {$ $ };
\node (50) at (4.5,5.) [draw,inner sep=7.1pt,label=left:Sr(38)] {$ $ };
\node (51) at (5.,3.5) [draw,inner sep=7.1pt,label=below:37] {$ $ };
\node (52) at (5,4.) [draw,inner sep=7.15pt] {$ $  };
\node (53) at (5.,4.5) [draw,inner sep=7.1pt] {$ $ };
\node (54) at (5.,5.) [draw,inner sep=7.1pt] {$   $ };
\node (55) at (5.,5.5) [draw,inner sep=7.1pt,label=left:Y(39)] {$  $ };
\node (56) at (5.5,4.) [draw,inner sep=7.1pt,label=below:38] {$ $ };
\node (57) at (5.5,4.5) [draw,inner sep=7.1pt] {$  $ };
\node (58) at (5.5,5) [fill,draw,inner sep=7.1pt] {$ $ };
\node (59) at (5.5,5.5) [draw,inner sep=7.1pt] {$ $ };
\node (60) at (5.5,6.) [draw,inner sep=7.1pt,label=left:Zr(40)] {$ $ };
\node (61) at (6.,4.5) [draw,inner sep=7.1pt,label=below:39] {$ $ };
\node (62) at (6.,5.) [draw,inner sep=7.1pt] {$ $ };
\node (63) at (6.,5.5) [draw,inner sep=7.1pt] {$ $ };
\node (64) at (6.,6.) [draw,inner sep=7.1pt] {$ $ };
\node (66) at (6.5,5.) [draw,inner sep=7.1pt,label=below:40] {$ $ };
\node (67) at (6.5,5.5) [draw,inner sep=7.1pt] {$ $ };
\node (68) at (6.5,6.) [fill,draw,inner sep=7.1pt] {$ $ };
\node (71) at (3.,1.) [draw,inner sep=7.1pt] {$ $ };

\draw [very thick]  (.5,0) to (.5,.5); 
\draw [very thick]  (.5,.5) to (.5,1.); 
\draw [very thick]  (.5,1.) to (1.0,.5); 
\draw [very thick]  (1.0,.5) to (1.0,1.); 
\draw [very thick]  (1.0,1.) to (1.5,.5); 
\draw [very thick]  (1.5,.5) to (1.5,1.); 
\draw [very thick]  (1.5,1) to (1.5,1.5); 
\draw [very thick]  (1.5,1.5) to (1.5,2.); 
\draw [very thick]  (1.5,2.) to (2.,1.5);
\draw [very thick]  (2.,1.5) to (2.,2.);
\draw [very thick]  (2.,2.) to (2.5,1.5); 
\draw [very thick]  (2.5,1.5) to (2.5,2.);
\draw [very thick]  (2.5,2.) to (2.5,2.5);
\draw [very thick]  (2.5,2.) to (3.0,1.5);
\draw [very thick]  (3.0,1.5) to (3.0,2.5);
\draw [very thick]  (2.5,2.5) to (2.5,3.);
\draw [very thick]  (2.5,3.) to (3.,2.5);
\draw [very thick]  (3.,2.5) to (3.,3.);
\draw [very thick]  (3.,3.) to (3.5,2.5);
\draw [very thick]  (3.5,2.5) to (3.5,3.);
\draw [very thick]  (4.,3.5) to (4.,4.);
\draw [very thick]  (4.,4.) to (4.5,3.5);
\draw [very thick]  (4.5,3.5) to (4.5,4.);
\draw [thick,gray]  (4.5,4.) to (4.5,4.5);
\draw [thick,gray]  (4.5,4.5) to (4.5,5.);
\draw [thick,gray]  (4.5,5) to (5,4.5);
\draw [very thick]  (5.,4.5) to (5,5.);
\draw [very thick]  (5,5.) to (5.5,4.5);
\draw [very thick] (5.5,4.5) to (5.5,5.);
\draw [thick,gray] (5.5,5) to (5.5,5.5);
\draw [thick,gray] (5.5,5.5) to (5.5,6.);
\draw [thick,gray] (5.5,6.) to (6,5.5);
\draw [very thick] (6,5.5) to (6.,6.);
\draw [very thick] (6.,6.) to (6.5,5.5);
\draw [very thick] (6.5,5.5) to (6.5,6.);
\draw [very thick] (4.5,4.) to (5,3.5);
\draw [very thick] (5,3.5) to (5,4.);
\draw [very thick] (5,4.) to (5,4.5);
\draw [very thick] (5,4.5) to (5,5.);
\draw [very thick] (5.5,5) to (6,4.5);
\draw [very thick] (6,4.5) to (6,5);
\draw [very thick] (6,5) to (6,5.5);
\draw [very thick] (6.,5.5) to (6,6.);
\draw [very thick] (3.5,3.) to (4.,2.5);
\draw [very thick] (4.,2.5) to (4.,3.);
\draw [very thick] (4.,3.) to (4.,3.5);

\end{tikzpicture}
\vskip -2cm

\caption{\label{fig:SRD} rp-process path for 1.2 GK (left panel) and 1.5 GK 
(right panel).}
\end{figure}
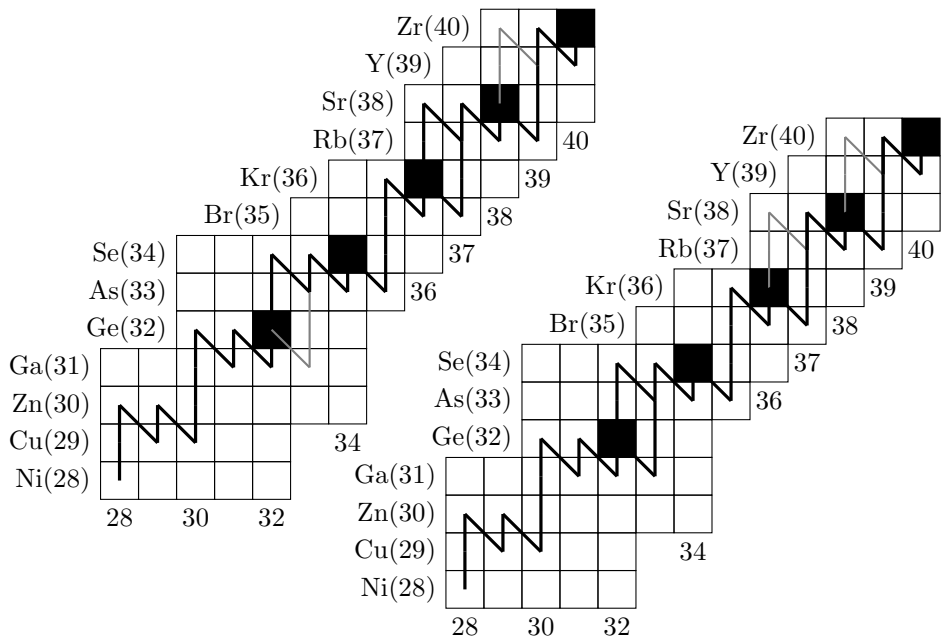 
In Fig. 3, we plot the path followed by the $rp$-process between $A=56$ and 
$A=80$ following the mass formula of Ref. I for $T=1.2$ GK and 1.5 GK. 
The dark boxes indicate the waiting points. The $Z$ values are indicated at the 
left of the diagram while the $N$ values are shown at the bottom. The lines 
indicate the path followed by nucleosynthesis. The paths through which 
more than 10\% of the total flux flows are indicated by black lines while gray 
lines show the corresponding paths for flux between 1\% - 10\%. 
The two waiting points, where Q-values  from the present formula have 
been used, are $^{64}$Ge and $^{76}$Sr. The former has been chosen as experimental values
differ widely. Experimental measurement is not available in $^{76}$Sr.
One can see that $^{64}$Ge is easily bridged by two proton capture at 1.2 GK. 
At 1.5 GK temperature, this waiting point delays the process so that 
$\beta$-decay of  $^{64}$Ge contributes significantly. At the next waiting 
point, $^{68}$Se, the photodisintegration is sufficiently strong so that,
independent of the temperature, $\beta$-decay is practically the only available 
path. This delays the nucleosynthesis significantly. At $^{72}$Kr, inverse 
process 
predominates in  higher temperature driving the flux through $\beta$-decay. 
Thus, here also, lower temperature helps nucleosynthesis speed up. 
The waiting point at $^{76}$Sr presents a different 
picture where the path essentially does not depend on the temperature and 
principally flows along decay. 
It is clear that the actual process is 
significantly dependent on the model of the burst process where the temperature 
and the proton fraction are functions of time. 
We find that at 1.2 GK, at the end of 100 seconds, the population that reaches 
$A=80$ or beyond is more than 1.5 times than the corresponding quantity at
the higher temperature of 1.5 GK. In both the cases, the population beyond
$A=76$ is significant.

\section{Summary}

The location of the proton dripline has been calculated using a new mass 
formula of Ref. I and compared with several other results as well as 
experiment. The new formula gives a good description of the dripline.
The implication of the mass prediction using the new formula on  $rp$-process
nucleosynthesis beyond $^{56}$Ni has been investigated. The masses predicted
by the present formula indicate
that nucleosynthesis proceeds easily up to mass 80 in a typical X-ray burst.

\section*{Acknowledgments}

The authors would like to thank one of the reviewers for pointing out 
the latest mass measurement of $^{65}$As.
This work has been carried out with financial assistance of the UGC sponsored
DRS Programme of the Department of Physics of the University of Calcutta.
Chirashree Lahiri acknowledges the grant of a fellowship awarded by the UGC.

\end{document}